\documentclass[12pt]{article}
\usepackage{amsmath,amssymb}
\newcommand{\beqn}{\begin{eqnarray}}
\newcommand{\eeqn}{\end{eqnarray}}
\newtheorem{thm}{Theorem}
\newtheorem{lem}{Lemma}

\newtheorem{defn}{Definition}

\newcommand{\proof}{\noindent {\it Proof.\/}\ }
\newcommand{\qed}{{\it Q.E.D.\/} \bigskip\par}
\newcommand{\defeq}{\stackrel{\mathrm{def}}{=}}
\newcommand{\Z}{{\mathbb{Z}}}
\newcommand{\C}{{\mathbb{C}}}
\renewcommand{\L}{\mathcal{L}}

\newcommand{\SL}{\mathop{{\mathrm{SL}}}}
\newcommand{\GL}{\mathop{{\mathrm{GL}}}}
\newcommand{\rd}{\partial}
\renewcommand{\O}{\mathcal{O}}
\renewcommand{\S}{\mathcal{S}}
\renewcommand{\H}{\mathcal{H}}
\newcommand{\tr}{\mathrm{tr}}
\renewcommand{\sl}{\mathfrak{sl}}
\newcommand{\g}{\mathfrak{g}}
\newcommand{\gnote}{\overset{\circ}{\g}}
\newcommand{\frakh}{\mathfrak{h}}
\renewcommand{\t}{\pmb{t}}
\newcommand{\z}{\pmb{z}}
\begin{document}
\title{Polynomial $\tau$-functions of the 
NLS-Toda hierarchy \\
and the Virasoro singular vectors}
\author{
Takeshi Ikeda\\
{\normalsize Department of Applied Mathematics,
Okayama University of Science}\\
{\normalsize Ridaicho 1-1, Okayama 700-0005, Japan}\\
{\normalsize E-mail: ike@xmath.ous.ac.jp}
\thanks{Partly supported by Grant-in-Aid for Scientific
Reseach (No.13740026) Japan Society for the Promotion of Science.}\\
Hiro-Fumi Yamada\\
{\normalsize Department of Mathematics, Okayama University}\\
{\normalsize Tsushima-naka 3-1-1, Okayama 700-8530, Japan}\\
{\normalsize E-mail: yamada@math.okayama-u.ac.jp}}
\date{}
\maketitle
\begin{center}
{\it To Minoru Wakimoto on his sixtieth birthday\/}
\end{center}

\bigskip

\begin{abstract}A family of polynomial $\tau$-functions for
the NLS-Toda hierarchy is constructed.
The hierarchy is associated with
the homogeneous vertex operator representation
of the affine algebra $\g$ of type $A_1^{(1)}.$
These $\tau$-functions are given explicitly
in terms of Schur functions that
correspond to rectangular Young diagrams.
It is shown that an arbitrary polynomial $\tau$-function
which is an eigenvector of $d$, the degree operator of $\g$,
is contained in the family.
By the construction, any $\tau$-function in the family
becomes a Virasoro singular vector.
This consideration gives rise to a simple
proof of known results on the Fock
representation of the Virasoro algebra with $c=1.$
\end{abstract}
\bigskip

Key words: Schur functions, nonliner Schr\"odinger equation, Virasoro algebra.

MSC 2000: 17B67,17B68,17B80,35Q55.

\newpage
\renewcommand{\theequation}{\arabic{section}.\arabic{equation}}
\section{Introduction}
\setcounter{equation}{0}

The main object of this paper is the family of the Schur
functions associated with the rectangular Young diagrams.
We have encountered these {\it rectangular Schur functions\/} in the 
studies of integrable systems and related representation
theories:
\begin{itemize}
\item
The Virasoro singular vectors in the Fock space of central charge $c=1$
\cite{bib:Segal},\cite{bib:WY},
\item
Polynomial $\tau$-functions of the NLS-Toda hierarchy \cite{bib:Sac},\cite{bib:GNF}.
\end{itemize} 
One of the aim of this paper is to understand these two results
in a unified manner. Another aim is to parametrize all the $\tau$-functions
of the NLS-Toda hierarchy which are homogeneous polynomials with respect 
to the degree specified below.
As the common background, we utilize
the homogeneous vertex operator construction of the basic representation
$L(\Lambda_0)$ of the affine Lie algebra $\g(A_1^{(1)})$ (\cite{bib:FK}).  

Among many ways of introducing the NLS-Toda hierarchy,
our main viewpoint is group-theoretic (\cite{bib:JM83},\cite{bib:KvL},\cite{bib:tKB}). 
One knows that the basic representation
$L(\Lambda_0)$ is realized on the space 
$V\defeq\C[\,\t\,] \otimes \C[e^\alpha,e^{-\alpha}],$
where $\t=(t_1,t_2,\ldots).$
The NLS-Toda hierarchy characterizes the $G$-orbit of the highest weight 
vector in $V$, where we denote by $G$ the group
associated with $\g(A_1^{(1)})$ introduced by Peterson and Kac \cite{bib:PK}.
The hierarchy is also described in terms of Hirota's 
bilinear forms. 
It is given by infinite number of 
bilinear equations including the following typical equations: 
\beqn
(D_{t_1}^2+D_{t_2})\tau_n\cdot \tau_{n+1}=0,\quad 
D_{t_1}^2\tau_{n}\cdot\tau_{n}+2\tau_{n-1}\tau_{n+1}=0\quad (n\in\Z)\label{eq:bin},
\eeqn
where we write a typical element $\tau$ of $V$
as $\tau=\sum_{n\in\Z}\tau_ne^{n\alpha}\;(\tau_n\in \C[\,\t\,]).$
The second equation of (\ref{eq:bin}) is known as the bilinear form
the Toda lattice equation.
If we set 
$\Phi\defeq\tau_{n+1}/\tau_n,\overline{\Phi}\defeq\tau_{n-1}/\tau_n$
for any fixed $n$ and pose the assumption
that $\overline{\Phi}$ is the complex conjugate of $\pm \Phi$,
then, from equations (\ref{eq:bin}), we have
\beqn
-i\rd_t\Phi=\rd_x^2\Phi\mp 2|\Phi|^2{\Phi},\label{eq:nls}
\eeqn
where $t\defeq  it_2$ and $x\defeq t_1.$
Equation (\ref{eq:nls}) is known as the nonlinear Schr\"odinger (NLS) 
equation. 
So we call this hierarchy as the NLS-Toda hierarchy.
Let us define the degree in the space $V$ by
\beqn
\mathrm{deg}(t_j)=j(j\geq 1),\quad
\mathrm{deg}(e^{n\alpha})=n^2(n\in \Z).
\eeqn
A direct computation of the $\mathrm{SL}_2(\C)$-orbit
of $e^{n\alpha}$ leads to a formula (\ref{eq:tau}) of  
$\tau$-functions of the hierarchy.
The $\tau$-functions are written in terms of rectangular Schur functions.
We can also prove that 
the family given by (\ref{eq:tau})
exhaust all the homogeneous $\tau$-functions of the hierarchy.

One knows that the Virasoro algebra acts on 
the space $V$, where the each sector
$\C[\,\t\,]e^{n\alpha}(n\in \Z)$ is
preserved.
It is remarkable that the Virasoro algebra
commutes with the Lie subalgebra $\mathfrak{sl}_2(\C)$ 
of $\g(A_1^{(1)}).$
Since our $\tau$-functions are given
as $\mathrm{SL}_2(\C)$-orbit through the vector $e^{n\alpha}$,
it is immediate to see that the $\tau$-function\
is a singular vector, that is we have $L_k \tau=0$
for $k>0.$ 
Hence we have a
simple explanation
of the result of Segal \cite{bib:Segal}
and Wakimoto-Yamada \cite{bib:WY}.

We here add some historical remarks. It should be firstly mentioned
that Sachs \cite{bib:Sac} obtained the formula for $\tau$-functions
in terms of rectangular Schur functions by using the Jacobi identity.
Gilson et al. \cite{bib:GNF} also derived the formula from the double
Wronskian type solution to the 2-component KP hierarchy.
In addition, M. Sato made the following inspiring comment,
unfortunately without any proof,
in his series of lectures delivered 
at Kyoto University,
: ``The whole family of the rectangular
Schur functions characterizes the NLS hierarchy''.

\section{Homogeneous construction of the basic representation}
First we recall some facts on the basic representation $L(\Lambda_0)$ of
the affine Lie algebra $\g$ of type $A_1^{(1)}$ due to
Frenkel and Kac (cf. \cite{bib:FK}, see also \cite{bib:Kac}).

Let $\gnote=\sl(2,\C)$ with the standard basis
\beqn
E=\left(\begin{array}{cc}0& 1\\0& 0\end{array}\right),\quad
H=\left(\begin{array}{cc}1& 0\\0& -1\end{array}\right),\quad
F=\left(\begin{array}{cc}0& 0\\1& 0\end{array}\right).\quad
\eeqn
The affine Lie algebra $\g$ of type $A_1^{(1)}$ can be realized
as the vector space
\beqn
\g={\gnote}\otimes \C[t,t^{-1}]\oplus \C K\oplus \C d
\eeqn
with the bracket
$[K,\g]=0,\;[d,X\otimes t^n]=nX\otimes t^n,$ and
\beqn
\left[X\otimes t^m,Y\otimes t^n\right]=\left[X,Y\right]
\otimes t^{m+n}+m\delta_{m+n,0}\tr(XY)K
\quad(X,Y\in \gnote,m,n\in\Z)
\eeqn
where $[X,Y]$ is the bracket in $\gnote.$
For $X\in \gnote, n\in\Z$ put $X_n\defeq X\otimes t^n.$
We identify $\gnote$
with the subalgebra of $\g$ consisting of the elements $X_0(X\in \gnote).$

Consider the subalgebra $\frakh=\C H\oplus \C K\oplus \C d.$
It is called a {\it Cartan subalgebra\/} of $\g.$
Let $\Lambda_0,\delta$ be the linear functions on $\frakh$ defined
by
\beqn
\Lambda_0(H)=0,\; \Lambda_0(K)=1,\;
\Lambda_0(d)=0,\;\delta(H)=\delta(K)=0,\;\delta(d)=1.
\eeqn
The {\it basic representation\/} $L(\Lambda_0)$ of the Lie algebra $\g$
is
the irreducible highest weight representation of the highest weight
$\Lambda_0.$
Namely it is an
irreducible $\g$-module in which there exists
a non-zero vector $v_{\Lambda_0}$
such that
\beqn
Kv_{\Lambda_0}=v_{\Lambda_0},\quad dv_{\Lambda_0}=0,\quad X_n
v_{\Lambda_0}=0
\label{eq:basic}
\;(n\ge 0).
\eeqn

We have
\beqn
[H_m,H_n]=2m\delta_{m+n,0}K.
\eeqn
The elements $H_n(n\ne 0)$ generate
a subalgebra $\H$, which is called the {\it homogeneous\/}
Heisenberg subalgebra of $\g.$
Let $\t$ denote the sequence of infinitely many independent variables
$t_1,t_2,\ldots.$
We have a natural representation of $\H$ on the space of
polynomials $\C[\,\t\,]$
given by
\beqn
H_n \mapsto 2\frac{\rd }{\rd t_n},\quad
H_{-n}\mapsto nt_n \;(n>0),\quad K\mapsto \mathrm{id}.\label{eq:actH}
\eeqn

We construct a representation
of $\g$ on the space
$V\defeq \C[\,\t\,]\otimes\C[Q]$,
where $\C[Q]$ is the group algebra of the root lattice $Q\defeq\Z\alpha$
of $\gnote.$
Let the elements of $\H$ act on the first factor
of $\C[\,\t\,]\otimes\C[Q]$ by (\ref{eq:actH}).
Define also an action of $H=H_0$ on $V$ by
\beqn
H(P\otimes e^{n\alpha})=2nP\otimes e^{n\alpha}\;(P\in \C[\,\t\,],n\in \Z).
\label{eq:actH0}
\eeqn
To describe the actions of $E_n$ and $F_n$, it is
convenient to consider the generating function
\beqn
X(z)\defeq\sum_{n\in\Z}X_n  z^{-n-1}\quad (X\in \gnote).
\eeqn
Let us introduce the following operators on $V$:
\beqn
z^{\pm H}(P\otimes e^{n\alpha})=z^{\pm 2n}P\otimes e^{n\alpha},\quad
e^{\pm\alpha}(P\otimes e^{n\alpha})=P\otimes e^{(n\pm 1)\alpha}\quad
(P\in\C[\,\t\,],\;n\in\Z).\label{eq:act0}
\eeqn
Then the action of $E(z)$ and $F(z)$
are given by the following {\it vertex operators\/}:
\beqn
E(z)\mapsto e^{\eta(\t,z)}e^{-2\eta(\widetilde{\rd}_{\t},z^{-1})}
e^{\alpha}z^{H},\quad
F(z)\mapsto e^{-\eta(\t,z)}e^{2\eta(\widetilde{\rd}_{\t},z^{-1})}
e^{-\alpha}z^{-H}\label{eq:actEF}
\eeqn
where
\beqn
\eta(\t,z)=\sum_{j=1}^\infty t_j z^j,\quad
\eta(\widetilde{\rd}_{\t},z^{-1})
=\sum_{j=1}^\infty \frac{1}{j} \frac{\rd}{\rd t_j}z^{-j}.
\eeqn
If we set $v_{\Lambda_0}\defeq 1\otimes e^0\in V$
then we have (\ref{eq:basic}).
It is shown that the set
of formulas (\ref{eq:actH}),(\ref{eq:actH0}),
(\ref{eq:act0}), and (\ref{eq:actEF}) gives
an irreducible representation of $\g$ on $V.$
In this way, we obtain a realization
of $L(\Lambda_0)$ on the space $V.$

\section{Definition of the NLS-Toda hierarchy}

Since $L(\Lambda_0)=V$ is an
{\it integrable\/} representation, we can define
the actions of $\exp(E_n),\exp(F_n)$ $(n\in\Z)$
on $V.$ These operators generate a
subgroup of $\GL(V)$ which we denote by $G.$
Let $\O=G v_{\Lambda_0}$ be the $G$-orbit through
the highest weight vector $v_{\Lambda_0}.$

\begin{defn}
  A non-zero vector $\tau\in V=\C[\t]\otimes \C[Q]$ is called
  a $\tau$-function of the NLS-Toda hierarchy
  if and only if $\tau\in \O.$
\end{defn}

Let $L_{high}$ be the highest component
of $V\otimes V$, i.e., the $\g$-submodule in $V\otimes V$ generated
by $ v_{\Lambda_0}\otimes v_{\Lambda_0}.$
Then we have $L_{high}\cong L(2\Lambda_0)$ and
the group orbit $\O$ can also be described as follows:
\beqn
\tau\in \O\Longleftrightarrow \tau\otimes \tau\in L_{high}.
\eeqn
Further, the method of the {\it generalized Casimir operator\/}
due to Kac and Wakimoto (\cite{bib:KW}) enables us to write down
the condition $\tau\otimes \tau\in L_{high}$ as
a system of Hirota's bilinear differential equations.
In particular, a $\tau$-function of the NLS-Toda hierarchy
satisfies (\ref{eq:bin}).
In this paper, however, we do not use the explicit forms of these
equations.
The bilinear equations are also derived 
from the free fermion operators and the boson-fermion 
correspondence (see \cite{bib:JM83},\cite{bib:KvL}).

\section{Rectangular Schur functions}\label{rec}

In this section we will show that the rectangular Schur functions appear naturally in 
the $\SL_2(\C)$-orbit through the maximal weight vector of $L(\Lambda_0)$.

Let the polynomials $p_k(\t)$ be defined by 
\beqn
e^{\eta(\t,z)}=\sum_{k=0}^\infty p_k(\t)z^k,
\eeqn
where $\eta(\t,z)=\sum_{j=1}^\infty t_j z^j$.
Then the Schur function $S_\lambda$ indexed by the partition $\lambda=(\lambda_1,\ldots,\lambda_n)$ can
be expressed by the following determinant:
\beqn
S_\lambda=S_\lambda(\t)=\det\left(p_{\lambda_i-i+j}(\t)\right)_{1\leq i,j \leq n},
\eeqn
where we agree that $p_k(\t)=0$ for $k<0$. 
Let $\lambda'$ be the conjugate of $\lambda.$
Then we have 
\beqn
S_\lambda(-\t)=(-1)^{|\lambda|}S_{\lambda'}(\t),\quad |\lambda|=\sum_{j=1}^n{\lambda_j}.\label{eq:transp}
\eeqn

If $\t$ is expressed by a sequence of finite number of variables 
$\z=(z_1,\ldots,z_n)$ as
\beqn
t_j=\frac{1}{j}\left(z_1^j+\cdots+z_n^j\right)\quad (j\geq 1)
\eeqn
then we have
\beqn
S_\lambda=S_\lambda(\z)=\frac{\det(z_i^{\lambda_{j}+n-j})_{1\leq i,j \leq n}}
{\det(z_i^{n-j})_{1\leq i,j \leq n} }.\label{eq:det/det}
\eeqn
The denominator
is nothing but the Vandermonde determinant:
\beqn
\Delta_n(\z):=\prod_{i<j}(z_i-z_j).
\eeqn
As for the rectangular Young diagram $(k,\ldots,k)$ (repeated $n$ times) denoted by 
$\square(n,k)$,
we see from (\ref{eq:det/det}) that $S_{\square(n,k)}(\z)$ has a simple form
\beqn
S_{\square(n,k)}(\z)=(z_1\cdots z_n)^k .
\eeqn

Let $\t=(t_1,t_2,\ldots)$ and $\z=(z_1,\ldots,z_n)$ be two sequences of 
infinite and finite variables, respectively.
The Cauchy formula (see \cite{bib:Mac}) can be written as
\beqn
e^{\sum_{k=1}^n \eta(\t,z_k)}=\sum_{\lambda}S_\lambda(\t)S_\lambda(\z),\label{eq:Cauchy}
\eeqn 
where summation runs over partitions $\lambda$ with length at most $n$.

We also note orthogonality relations for the $S_\lambda(\z).$
If $f=f(z_1,\ldots,z_n)$ is a Laurent polynomial, let 
$\overline{f}=f(z_1^{-1},\ldots,z_n^{-1})$ and 
let $CT[f]$ denote the constant term in $f.$
A scalar product is defined by
$
\langle f,g
\rangle_n=\frac{1}{n!}CT[f\overline{g}\Delta_n\overline{\Delta_n}].
$
Then we have
\beqn
\langle S_\lambda,S_\mu\rangle_n=\delta_{\lambda\mu}.\label{eq:ortho}
\eeqn

Our main result in this section is the following.

\begin{thm}\label{thm:tau}
Let $k$ be a non-negative integer, and $\gamma\in\C$.
We have the following expression of 
a $\tau$-function:
\beqn
\exp({\gamma F})\left(e^{k\alpha}\right)
=\sum_{n=0}^{2k}
(-1)^{\frac{n(n+1)}{2}}
\gamma^n
S_{\square(2k-n,n)}(\t)e^{(k-n)\alpha}.\label{eq:tau}
\eeqn
\end{thm}
\proof
First we note that 
\beqn
F(z_1)\cdots F(z_n)e^{k\alpha}
=e^{-\sum_{j=1}^n\eta(\t,z_j)}(z_1\cdots z_n)^{-2k}\Delta_n^2\, e^{(k-n)\alpha}.\label{eq:Fe}
\eeqn
It can be verified by using the relations
\beqn
e^{-\eta(\t,z)}e^{-2\eta(\widetilde{\rd_{\t}},w^{-1})}
=\left(1-\frac{w}{z}\right)^2e^{-2\eta(\widetilde{\rd_{\t}},w^{-1})}e^{-\eta(\t,z)},\quad
z^{-H}q^{-1}=z^2 q^{-1}z^{-H}.
\eeqn
Now we want to pick up the coefficient of $(z_1\cdots z_n)^{-1}$ in (\ref{eq:Fe}), which is equal to $F^n e^{k\alpha}.$
Notice that the coefficient can be written in terms of the scalar product
\beqn
F^n e^{k\alpha}
&=&CT[z_1\cdots z_n F(z_1)\cdots F(z_n)e^{k\alpha}]\nonumber\\
&=&\sum_{\lambda}(-1)^{\frac{n(n-1)}{2}}
S_\lambda(-\t)
n! \langle S_\lambda,{S_{\square(n,2k-n)}}\rangle_n e^{(k-n)\alpha},
\eeqn
where we expand
$e^{-\sum_{j=1}^n\eta(\t,z_j)}$
by (\ref{eq:Cauchy}) and 
use $\overline{\Delta_n}=(-1)^{\frac{n(n-1)}{2}}(z_1\cdots z_n)^{-n+1}\Delta_n.$
Using the orthogonality (\ref{eq:ortho}) and (\ref{eq:transp}) we have
\beqn
F^n e^{k\alpha}=n!(-1)^{\frac{n(n-1)}{2}}S_{\square(n,2k-n)}(-\t)
=n!(-1)^{\frac{n(n+1)}{2}}S_{\square(2k-n,n)}(\t).\label{eq:F^ne^k}
\eeqn
\qed

\section
{Parametrization of the homogeneous $\tau$-functions}

\begin{defn}
  If $\tau\in \O$ is an eigenvector of $d$ with eivenvalue $N$
  then we say that $\tau$ is homogeneous of degree $N.$
\end{defn}

We first state the following theorem that claims
the family of $\tau$-functions constructed in Section \ref{rec}
exhausts all the homogeneous polynomial $\tau$-functions.

\begin{thm}\label{thm:param}
  Let $\tau$ be a homogeneous $\tau$-function of degree $N.$
  Then there exists a non-negative integer
  $m$ such that $N=m^2.$
  If, moreover, $\tau$ is not constant multiple of $e^{\pm m\alpha}$, then
  there exist constants
  $\gamma,c\ne 0$ such that $\tau=c\exp\left({\gamma F}\right)e^{m\alpha}.$
\end{thm}

To prove Theorem \ref{thm:param}, we will use a
fundamental result by Peterson-Kac.
Let $\g$ be a
symmetrizable Kac-Moody Lie algebra, and $\frakh$ be its Cartan
subalgebra.
Let $\Lambda\in \frakh^*$ be a dominant integral weight
and $L(\Lambda)$ be an irreducible $\g$-module
with highest weight $\Lambda$.
We denote by $v_{\Lambda}$ a highest weight vector
in $L(\Lambda)$.
For $v\in L(\Lambda)$ let
$v=\sum_{\lambda\in \frakh^*}v_\lambda$
be its weight decomposition.
Put ${\mathrm{supp}}(v)\defeq\{\lambda\in P(\Lambda)\;|\;
v_{\lambda}\ne 0\}$, and let $S(v)$ be the convex hull of
${\mathrm{supp}}(v)$ in $\frakh^*$.
By $G\subset \GL(L(\Lambda))$ we denote the associated group.
And let $W$ denote the Weyl group of $\g.$

\begin{lem}\cite{bib:PK}\label{lem:PK}
  Let $\tau\in L(\Lambda)$ be in the $G$-orbit through $v_{\Lambda}$.
  Then each vertex of $S(\tau)$
  belongs to the $W$-orbit
  of the highest weight $\Lambda$.
\end{lem}

\bigskip
\noindent {\it Proof of Theorem \ref{thm:param}.\/}
For $\g=\g(A_1^{(1)})$, $W$-orbit of $\Lambda_0$
consists of the elements
\beqn
\lambda_m\defeq \Lambda_0+m\alpha-m^2\delta\;(m\in\Z).
\eeqn
Note that $\lambda_m$ is the weight of $e^{m\alpha}\in V.$
Let $\tau\in {\cal{O}}$ be homogeneous of degree $N.$
By applying Lemma \ref{lem:PK},
we see that there exists a non-negative integer $m$ such
that $N=m^2$ and that
$S(\tau)$ is the segment
\beqn
[\lambda_m,\lambda_{-m}]\defeq
\{s\lambda_m+(1-s)\lambda_{-m}\;|\;0\le s\le 1\}
\eeqn
unless $S(\tau)=\{\lambda_m\}$ or $\{\lambda_{-m}\}.$
If $S(\tau)=\{\lambda_{\pm m}\}$ then
$\tau$ is a constant multiple of $e^{\pm m\alpha}$
respectively.
Therefore we assume $S(\tau)=[\lambda_m,\lambda_{-m}]$
and write
$
\tau=\sum_{n=0}^{2m}\tau_{m-n} e^{(m-n)\alpha},
$
where we have $\deg\,\tau_n=n(2m-n)$.
In particular,
$\tau_{\pm m}$ is a non-zero constant.

For a complex parameter $\gamma$, we set
$\sigma\defeq\exp(-\gamma F)\tau.$
By the definition of $\O$, we have $\sigma\in {\cal{O}}.$
In addition, since $F$ preserves the degree,
we can write $\sigma
=\sum_{n=0}^{2m}\sigma_{m-n} e^{(m-n)\alpha}$,
$\deg \sigma_n=n(2m-n)$, in a similar way to $\tau$.
Now we see the coefficient
$\sigma_{-m}$ is constant in $\t$, and
more explicitly we have
\beqn
\sigma_{-m}=(-1)^{m}\tau_m \gamma^{2m}+\mbox{lower order term in}\;\gamma.
\eeqn
by (\ref{eq:F^ne^k}).
Now we can choose $\gamma$ such that the polynomial $\sigma_{-m}$
is equal to zero. Thus $\lambda_{-m}\not\in \mathrm{supp}(\sigma).$
Then by Lemma \ref{lem:PK}, we have $S(\sigma)$
consists of the one point $\lambda_m.$
That is to say $\sigma=ce^{m\alpha}$
for certain non-zero constant $c\in\C$.
Hence we have $\tau=c\exp(\gamma F)e^{m\alpha}$.
\qed

\section{Virasoro singular vectors}
Let define the operators $L_n\;(n\in \Z)$ on $V$ by
\beqn
L_n=\frac{1}{4}\sum_{m\in\Z}:H_{m+n}H_n:,
\eeqn
where we define the {\it normal ordering\/} by 
\beqn
:H_mH_n:=\begin{cases}
H_mH_n\;\mbox{if}\;m\le n\\
H_nH_m\;\mbox{if}\; m>n.
\end{cases}
\eeqn
Then we have
\beqn
\left[L_m,X_n\right]&=&-nX_{m+n},\label{eq:sug}\\
\left[L_m,L_n\right]&=&(m-n)L_{m+n}+\frac{1}{12}(m^3-m)\delta_{m+n,0}\;\mathrm{id}.\label{eq:vir}
\eeqn
Equation (\ref{eq:vir}) means that the operators $L_n$ 
gives a representation of the Virasoro algebra $\L$ with
central charge $c=1.$
For $N\in \Z$, we set
\beqn
\S(N)\defeq \{
v\in V\,|\,L_k v=0(k\geq 1),\; L_0 v=N v
\}.\quad
\eeqn
We also set $\S_n(N)\defeq \S(N)\cap V_n\;(n\in \Z).$ Then we have $\S(N)=\bigoplus_{n\in \Z} \S_n(N).$
A vector $v$ in $\S_n(N)$ is called a {\it singular vector\/} in $V_n$ 
of {\it grade\/} $N.$
For any $n\in\Z$, $e^{n\alpha}\in V_m$ is a singular vector of grade $n^2.$
This follows from $H_k e^{n\alpha}=0$ for $k>0$ and $He^{n\alpha}=2ne^{n\alpha}.$ 

Now we remark that equation (\ref{eq:sug}) implies, in particular, the following notable fact:
\beqn
[\gnote,\L]=0\quad \mbox{on}\; V.
\eeqn
Therefore, starting from a singular vector $e^{m\alpha}$, $\gnote$ creates other singular
vectors . By virtue of Theorem \ref{thm:tau}, they are nothing but rectangular 
Schur functions. 

\begin{thm}(\cite{bib:Segal},\cite{bib:WY}) If there exists a non-negative integer $m$ such that
$N=m^2$
then
\beqn
\S(N)=\bigoplus_{-m\le n\le m}
\C S_{\square(m-n,m+n)}(\t)e^{n\alpha},\label{eq:S(N)}
\eeqn
and then $\S(N)$ is an irreducible $\gnote$-module,
otherwise $\S(N)=\{0\}.$
\end{thm}
\proof
Let $m$ be a non-negative integer. 
Put $N=m^2.$ 
By the discussion above, we have
\beqn
S_{\square(m-n,m+n)}(\t)e^{n\alpha}\in \S_n(N),
\eeqn
for $n\in \Z$ such that $-m\le n\le m.$
The irreducibility of the right hand side of (\ref{eq:S(N)}) as $\gnote$-module can 
be seen from its character.
Let $W_{m,n}\subset V$ be the $\L$-submodule generated by $S_{\square(m-n,m+n)}(\t)e^{n\alpha}.$
Then $W_{m,n}$ is an irreducible highest weight $\L$-module.
In fact, $V$ is known to be completely reducible because it is {\it unitarizable.}
Now we can make use of the following character formula due to Kac (\cite{bib:Contra}): 
\beqn
\tr_{W_{m,n}} q^{L_0}
=\frac{q^{m^2}(1-q^{2m+1})}{\prod_{k=1}^\infty (1-q^k)}.
\eeqn
Here we set $W=\bigoplus_{m=0}^\infty \bigoplus_{-m\le n\le m} W_{m,n}.$
To complete the proof, it is suffices to note $W=V$, which
is due to Frenkel \cite{bib:F}.
Namely, we have the following character identity:
\beqn
\tr_{W} z^H q^{L_0}&=&\sum_{m=0}^\infty\sum_{-m\le n\le m}\tr_{W_{m,n}}z^H q^{L_0}\nonumber\\
&=&\sum_{m=0}^\infty\sum_{-m\le n\le m}
\frac{z^{2n}q^{m^2}(1-q^{2m+1})}{\prod_{k=1}^\infty (1-q^k)}\nonumber\\
&=&\frac{\sum_{n\in\Z} z^{2n}q^{n^2}}{\prod_{k=1}^\infty (1-q^k)}=\tr_{V} z^H q^{L_0}.
\eeqn
\qed
The irreducible decomposition of the Virasoro module $V_n$ and
the formulas of the singular vectors in terms of rectangular Schur functions
are wellknown. These results were obtained more than 15 years ago by Segal \cite{bib:Segal}
and, independently, by Wakimoto-Yamada \cite{bib:WY}.
In the present paper, we make use of the action of the affine Lie algebra on the
space  $V=\oplus_{n\in\Z}V_n$.
This viewpoint enables us to explain the formula of singular vectors
quite naturally.

\section{Concluding remarks}
Taking each of the maximal weight vectors $e^{m\alpha}$ in $L(\Lambda_0)$
as a {\it seed\/} solution,
we have constructed all of the  $\tau$-function of the NLS-Toda hierarchy,
which are eigenvectors of $L_0.$
It turns out that each of them constitutes
a finite Toda chain of the Schur functions associated with rectangular
Young diagrams.
Moreover, we have got a clear understanding
that each component $\tau_me^{m\alpha}$ of a $\tau$-function 
is a singular vector of the Virasoro algebra.
It will be an interesting problem to describe
the $\tau$-functions with a support on the arbitrary polyhedra other than
the segment $[\lambda_{-m},\lambda_m].$


\bigskip

\begin{thebibliography}{99}

\bibitem{bib:Segal}
Segal, G.: Unitary representations of some infinite-dimensional groups,
{\it Comm. Math. Phys.} {\bf 80}, no. 3 (1981), 301--342. 

\bibitem{bib:WY}
Wakimoto, M. and Yamada, H.: The Fock representations of the Virasoro algebra and 
the Hirota equations of the modified KP hierarchies,
{\it Hiroshima Math. J.\/} {\bf 16}, no. 2 (1986), 427--441.

\bibitem{bib:Sac}
Sachs, R. L.: Polynomial $\tau$-functions for the AKNS hierarchy,
in: {\it Theta functions---Bowdoin 1987, Part 1}, Amer. Math. Soc., Providence, RI,
1989, {\it Proc. Sympos. Pure Math.} {\bf 49}, Part 1 , pp.133--141

\bibitem{bib:GNF}
Gilson, C. R., Nimmo, J. J. C. and Freeman, N. C.: Rational solutions to the 
two-component K-P hierarchies,
in: {\it Nonlinear evolution equations and dynamical systems}, Springer, Berlin, 1990, pp. 32--35

\bibitem{bib:FK}
Frenkel, I. B.  and Kac, V. G.: Basic representations of affine Lie algebras 
and dual resonance models,
{\it Invent. Math.} {\bf 62}, no. 1 (1980/81), 23--66.

\bibitem{bib:JM83}
Jimbo, M. and Miwa, T.: Solitons and infinite-dimensional Lie algebras,
{\it Publ. Res. Inst. Math. Sci.} {\bf 19}, no. 3 (1983), 943--1001.

\bibitem{bib:tKB}
ten Kroode, A. P. E. and Bergvelt, M. J.: The homogeneous realization of 
the basic representation
of $A\sp {(1)}\sb 1$ and the Toda lattice,
{\it Lett. Math. Phys.}
{\bf 12}, no. 2 (1986), 139--147.

\bibitem{bib:KvL}
Kac, V. G. and van de Leur, J. W.: The $n$-component KP hierarchy and 
representation theory,
in {\it Important developments in soliton theory\/},
Springer Ser. Nonlinear Dynam., Springer, Berlin (1993) 302--343

\bibitem{bib:PK}
Peterson, D. H. and Kac, V. G.: Infinite flag varieties and conjugacy theorems,
{\it Proc. Natl. Acad. Sci. USA\/}
{\bf 80}, no. 6i. (1983), 1778--1782.

\bibitem{bib:Kac}
Kac, V. G.: {\it Infinite-dimensional Lie algebras\/}, Third edition. 
Cambridge University Press, Cambridge, 1990.

\bibitem{bib:KW}
Kac, V. G. and Wakimoto, M.: Exceptional hierarchies of soliton equations,
in: {\it Theta functions---Bowdoin 1987, Part 1},
Amer. Math. Soc., Providence, RI, 1989,
{\it Proc. Sympos. Pure Math.} {\bf 49}, Part 1, pp.191--237 


\bibitem{bib:Mac}
Macdonald, I. G.: {\it Symmetric functions and Hall polynomials\/},
Second edition. With contributions by A. Zelevinsky.
Oxford Mathematical Monographs. Oxford Science Publications.
The Clarendon Press, Oxford University Press, New York, 1995.


\bibitem{bib:Contra}
Kac, V. G.: Contravariant form for the infinite-dimensional Lie algebras
and superalgebras, 
{\it Lecture Notes in Phys.} {\bf 94} (1979), 441--445.


\bibitem{bib:F}
Frenkel, I. B.: Representations of affine Lie algebras,
Hecke modular forms and Korteweg-de
Vries type equations,
in: {\it Lie algebras and related topics\/}
(New Brunswick, N.J., 1981), pp. 71--110, Lecture Notes in
Math. {\bf 933}, Springer, Berlin-New York, 1982.


\end{thebibliography}
\end{document}